\documentclass[aps,prb,floatfix,twocolumn,showpacs]{revtex4}
\usepackage{graphicx}

\begin{document}

\title{Local spin valve effect in lateral (Ga,Mn)As/GaAs spin Esaki diode devices}

\author{M. Ciorga, C. Wolf, A. Einwanger, M. Utz, D. Schuh, and D. Weiss}
\affiliation{Experimentelle und Angewandte
Physik, University of Regensburg, D-93040 Regensburg, Germany.}

\begin{abstract}
We report here on a local spin valve effect observed unambiguously in lateral all-semiconductor all-electrical spin injection devices, employing $p^{+}-$(Ga,Mn)As/$n^{{+}}-$GaAs Esaki diode structures as spin aligning contacts. We discuss the observed local spin-valve signal as a result of interplay between spin-transport-related contribution and tunneling anisotropic magnetoresistance of magnetic contacts. The magnitude of the spin-related magnetoresistance change is equal to 30$\Omega$ which is twice the magnitude of the measured non-local signal. 
\end{abstract}


\maketitle

There has been recently a big progress on all-electrical spin injection and detection in lateral semiconductor devices with GaAs-based\cite{lou,ciorga2009} and Si transport channels.\cite{Jansen} Most of the reported experiments were performed on devices operating in a $non-local$ (NL) configuration, in which spin accumulation generated in the transport channel is probed by a detector contact placed within a certain distance from the injector, outside the current path. The electrical signal which is detected in such a configuration is then a measure of a pure spin current flowing beneath a detector.\cite{johnson,jedema} Whereas this technique proved to be powerful at studying problems related to electrical spin injection and detection, it may not be sufficient for employing in operational spintronics devices, e.g. spin FET.\cite{spinFET} A prerequisite for several concepts of a spin transistor is the electrical spin signal in $local$ configuration, i.e., with spin-polarized charge current flowing between spin-polarized source and drain contacts. The measure of a spin signal is then a relative magnetoresistance change $\Delta R/R_{P}$ where $\Delta R=R_{AP}-R_{P}$ and $R_{P(AP)}$ is resistance measured in parallel (antiparallel) configuration of magnetizations in source and drain contacts. The conditions required for observation of an efficient electrical spin signal were discussed extensively in some theoretical papers.\cite{Fert_Jaffres,Fert,Dery} According to those studies the crucial parameter governing the efficiency is the contact tunnel resistance $R^{\ast}_{b}$ at the interface between ferromagnetic material and semiconductor, or, speaking more precisely, the ratio of $R^{\ast}_{b}$ and the product $r_{N}=\rho_{N}\lambda_{N}$, where $\rho_{N}, \lambda_{N}$ are the resistivity and spin diffusion length of the non-magnetic semiconducting material, respectively. High value of parameter $R^{\ast}_{b}/r_{N}$ enables efficient spin injection overcoming so-called $conductivity\ mismatch$\cite{Schmidt} between ferromagnet and non-magnetic material. A too high ratio $R^{\ast}_{b}/r_{N}$ makes spins relax before they can be detected, preventing this way an efficient electrical detection of the signal. As a result there exists a window in the possible values of the parameter $R^{\ast}_{b}/r_{N}$ for which the obtained electrical spin signal is optimal.\cite{Fert} The above applies both for non-local and local measurements, however in the latter the measured spin signal must compete with magnetoresistance effects at the source and drain contacts, making actual detection of the signal difficult. As $R^{\ast}_{b}/r_{N}$ is usually pretty high for metal/semiconductor interfaces it brought Fert $et\ al.$\cite{Fert} to conclusion that transport channels from other than semiconducting materials, e.g carbon nanotubes, could be much more suitable for spin-FET type of devices.

\begin{figure}
\label{f1}
\begin{center}
\includegraphics[width=0.8\columnwidth,clip]{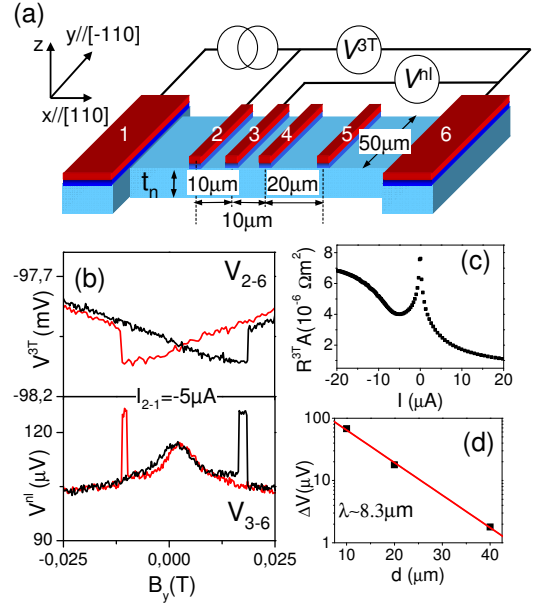}
\caption{(color online) (a) schematics of the experimental device; (b) typical dependence of the non-local (NL) voltage (bottom) and three-terminal (3T) voltage (top) on in-plane magnetic field. Measurement configuration as shown in (a);(c) bias dependence of resistance-area product  for contact 2 at B=0, typical for all used contacts. (d) dependence of NLSV signal on injector--detector separation}
\end{center}
\end{figure}

In this paper we report on experiments with all-semiconductor lateral devices employing $p^{+}-$(Ga,Mn)As/$n^{{+}}$-GaAs Esaki diode source and drain contacts. In our previous work\cite{ciorga2009,ciorga2009b} we reported on a successful implementation of an efficient all-electrical spin injection and detection scheme in such a system. Esaki diode structure in the contacts ensures that under small applied bias electrons can tunnel between the valence band of (Ga,Mn)As and the conduction band of GaAs.\cite{kohda2001,johnston,vandorpe2004} We have measured a relatively high spin injection efficiency $P$ of $\approx 50\%$ for low bias currents $|I|\leq 10\mu A$. The ratio $R^{*}_{b}/r_n\approx100$ for investigated devices places this value on the edge of the local spin valve effect observability window.\cite{Fert} The fact, that in those devices we could not obtain parallel-antiparallel configuration in magnetic contacts made measurements in local configuration even more difficult. Here we present measurements on a similar device, however with a slightly different geometry. We clearly observe different switching fields for employed source and drain contacts and a clear spin valve signal in a local configuration. The amplitude of the signal being 0.1$\%$ is certainly not optimal yet but consistent with predictions of Ref.~\onlinecite{Fert}. We discuss in the end that it should be feasible to optimize the parameter $R^{*}_{b}/r_n$ for this type of devices to the value close to the optimal.   

\begin{figure}
\label{f2}
\begin{center}
\includegraphics[width=0.9\columnwidth,clip]{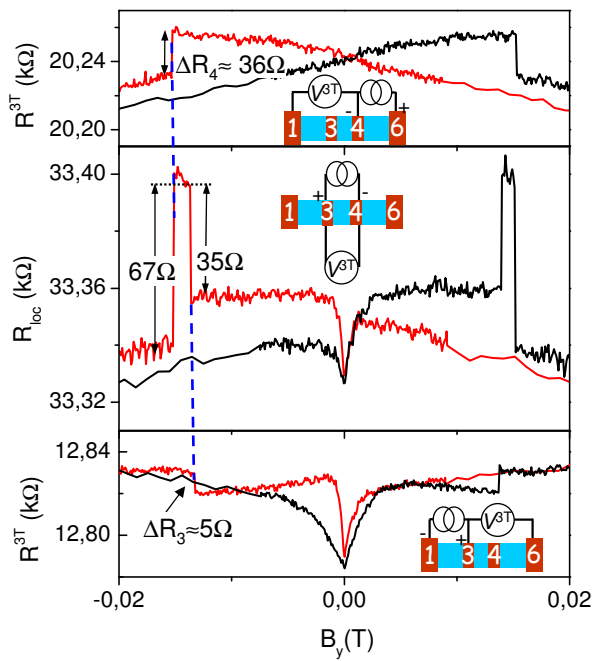}
\caption{(color online) Local magnetoresistance curve measured between contacts 3 and 4 (middle panel) vs. 3T magnetoresistance curves of the corresponding idividual contacts. Measurements performed for injection current of $\pm5\mu A$. Measurements configurations are shown as insets.}
\end{center}
\end{figure}

The experiments were performed on devices of a similar type as the one used in Ref.~\onlinecite{ciorga2009}. The schematic of the device is shown in Fig. 1(a). The sample features six magnetic Esaki diode contacts to the transport channel. Four contacts in the middle (2--5) are used to inject or detect spins in the channel. The size of those contacts is ($4\times50 \mu m^2$) and the spacing between their centers is 10 $\mu$m between pairs 3--2, 4--3 and 20 $\mu$m between the pair 5--4. Two outside contacts (1, 6), placed around $300\mu m$ from the center area, are much bigger ($150\times150 \mu m^2$) and are used as reference contacts in non-local measurements. The Esaki diodes consist of 50 nm of Ga$_{0.95}$Mn$_{0.05}$As and 8 nm of $n^{+}$-GaAs, with $n^{+}=6\times10^{18} $cm$^{-3}$. The transport channel is a $1\mu m$ thick $n-$GaAs layer with $n=2.7\times10^{16} $cm$^{-3}$. Between the diode and the channel a 15 nm thick GaAs transition $n^{+}\rightarrow n$ layer is also used. The mesa is oriented along [110] and contacts along $[1\bar{1}0]$ direction. In Fig. 1(b) one can see typical results of non-local spin-valve (NLSV) measurements  with spins injected at the contact 2 and external magnetic field applied along $[1\bar{1}0]$. NLSV signal observed at the detector 3 is shown in the bottom panel whereas  in the upper one we show a three-terminal (3T) voltage $V_{2-6}$, which is a measure of magnetoresistance of the interface.\cite{ciorga2009b} From the dependence of the amplitude of the NLSV signal on injector--detector separation, shown in (d), we estimate the spin diffusion length in the channel as $8.3 \mu m$.  Bias dependence of the product of the resisistance and area of the injector contact, which is a measure of $R^{\ast}_{b}$, is shown in Fig. 1(c). We can see that the dependence is quite asymmetric with lower value for positive bias values than for negative ones. Given the measured resistivity of the channel $\rho_{N}=1.3\times 10^{-3}\Omega m$ and the spin diffusion length $\lambda_{N}=8.3\mu m$ we get $R^{\ast}_{b}/r_{N}\approx 2-4\times10^{2}$ for $I=\pm 5\mu A$ (the current used in further measurements). The $R^{\ast}_{b}/r_{N}$ value is then even slightly bigger then for our previous sample. The switching behavior of the contacts is however different. We get information on the latter from 3T measurements thanks to the tunneling anisotropic magnetoresistance (TAMR) effect\cite{gould} at the Esaki diode contacts.\cite{ciorga2007} One can clearly see that switching fields observed in $V_{2-6}$ correspond to the higher switching field value observed in NL signal of $V_{3-6}$. The lower switching fields we thus attribute to the detector contact 3.

In the middle panel of Fig. 2 we show the results of local spin valve (LSV) measurements involving contacts 3 and 4, separated by 10 $\mu m$. The resistance in local configuration $R^{loc}$ is the sum of the channel resistance and the resistance of the individual contacts. Any MR effects observed in $R^{loc}$ are then a superposition of effects observed in $R^{3T}$ and the investigated  spin transport effects in the channel. We compare then the local measurements with the measurements performed on contact 4 (top panel) and contact 3 (bottom). One can clearly see that the switching fields observed in MR traces of the individual contacts match very well the switching fields observed in LSV measurements. When we subtract the resistance jumps observed in those 3T traces (36$\Omega$ and 5$\Omega$ for contact 4 and 3, respectively) from the resistance jumps observed in LSV signal we obtain the amplitude $\Delta R^{loc}=30\Omega$, which is then the amplitude of the local magnetoresistance change due to change of magnetization configuration in source and drain contacts from parallel to antiparallel. This gives us relative change $\Delta R/R\approx 0.1\%$, which is consistent with findings of Ref.~\onlinecite{Fert}, given the value of $R^{\ast}_{b}/r_{N}$ in our devices.

\begin{figure}
\label{f3}
\begin{center}
\includegraphics[width=0.9\columnwidth,clip]{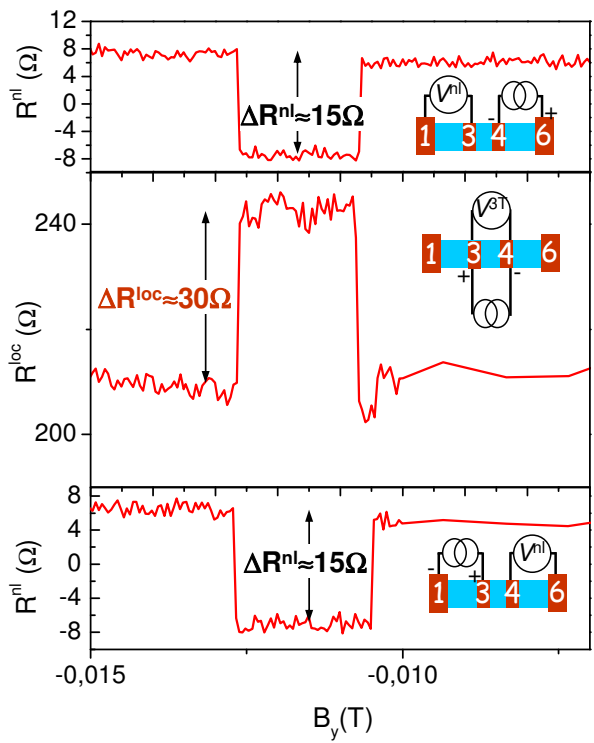}
\caption{(color online) Local SV signal (middle panel) vs. NLSV signals (top and bottom panels). LSV curve was obtained by subtracting 3T resistance of both involved contacts ($R^{3T}_{3}$ and $R^{3T}_{4}$) from the local resistance $R^{loc}_{3,4}$ measured between contacts 3 and 4. Measurements were performed for injection current of $\pm5\mu A$. Measurement configurations are shown as insets. }
\end{center}
\end{figure}

To check further that the measured $\Delta R$ is indeed due to spin-polarized transport we compare LSV and NLSV measurements in Fig. 3. The former is shown in the middle panel. For clarity we show only the results of a down-field sweep in the range around the SV feature. The plotted local data are obtained by subtracting MR traces of individual contacts from the local measurements. As a result we obtain a curve with a clear SV signal with the amplitude $\Delta R=30\Omega$ as discussed in the previous paragraph. In top and bottom panels we show the NL resistance curves measured between the contacts 3--1 and 4--6, with the current $I_{4-6}=-5\mu A$ and $I_{3-1}=5\mu A$, respectively. We see that switching field values in NLSV signal match very well those observed for LSV signal. The amplitude of NLSV $\Delta R^{nl}$ is in both cases around $15\Omega$, i.e., $\Delta R^{loc}=2\Delta R^{nl}$ what is expected from theory\cite{jedema2003} and what was also observed in graphene-based devices.\cite{weihan} This confirms that the spin-valve-like signal observed in a local configuration is indeed due to spin-polarized transport.

Let us now discuss shortly the possibilities of improving some of device parameters in order to increase the amplitude of LSV signal. To lower the value of $R^{\ast}_{b}/r_{N}$ one needs, of course, either to lower $R^{\ast}_{b}$ or increase $r_{N}=\rho_{N}\lambda_{N}$. The latter should be done easily by lowering the doping that would increase both resistivity $\rho_{N}$ and spin diffusion length $\lambda_{N}$. This is what we did in current devices in comparison to those investigated in Ref.~\onlinecite{ciorga2009} and as a result we increased $r_{N}$ from $5\times10^{-10}\Omega m^{2}$ to $1\times10^{-8}\Omega m^{2}$. Unfortunately $R^{\ast}$ increased by roughly the same factor. We would like to point out here however, that we checked many devices from the same wafer material and some of them showed $R^{\ast}_{b}\approx3\times10^{-8}\Omega m^{2}$, i.e., even slightly lower than in those other samples.\cite{comment1} This suggests either non-uniformity in wafer parameters or that the fabrication process could affect the actual interface resistance. Further work on the subject would have to involve finding the way to understand and control those effects to keep $R^{\ast}_{b}$ value as low as possible and bring the values for LSV above 10\%. This would be very reasonable number in terms of application in possible devices.

In summary, we have demonstrated unambiguous observation of a local spin valve effect in lateral all-semiconductor spin injection devices. Although the absolute amplitude of the signal is not very big, our experiments show that optimizing some parameters of our devices, namely interface resistance, and increasing the amplitude of local spin valve signal is feasible for this type of devices.

This work has been supported by Deutscheforschungsgemeinschaft (DFG) through SFB689 project.



\begin{thebibliography}{00}
\bibitem{comment1}
Because of a geometry of the contacts we were not able to obtain parallel-antiparallel configuration of contacts magnetizations in those devices.
\end{thebibliography}
\end{document}